\documentclass[twocolumn,showpacs,preprintnumbers,amsmath,amssymb]{revtex4}
\usepackage{graphicx}% Include figure files
\usepackage{dcolumn}% Align table columns on decimal point
\usepackage{bm}% bold math

\begin{document}

%\preprint{APS/123-QED}

\title{Neutrino clustering in the galaxy with a global monopole}
%Manuscript Title:\\with Forced Linebreak}% Force line breaks with \\

\author{Tae Hoon Lee}%
\altaffiliation[Permanent address: ]{  Department of Physcis, Soongsil University, Seoul 156-743, Korea; thlee@physics.ssu.ac.kr}
\author{Bruce H. J. McKellar  } 
\email 
{ b.mckellar@physics.unimelb.edu.au}
%Physics Department, XYZ University.}%Lines break automatically or can be forced with \\

\affiliation{%
School of Physics, University of Melbourne, Victoria 3010, Australia}

\date{\today}% It is always \today, today,

\begin{abstract}
In  spherically symmetric, static spacetime, we show that only $j=
1/2$ fermions can satisfy both Einstein's field equation and Dirac's
equation.  It is also shown that neutrinos are able to have effective
masses and cluster in the galactic halo when they are coupled to a
global monopole situated at the galactic core.  Astronomical
implications of the results are discussed.
\end{abstract}

\maketitle

\section{Introduction}	
%) A SECTION HEADING

In Barriola and Vilenkin's global monopole solution \cite{Vil} to
Einstein's field equation, scalar fields with global $O(3)$ broken
symmetry are minimally coupled to gravity and the background spacetime
has deficits of angle.  Nucamendi {\it et al.} \cite{Nuc} suggested
that the global monopole solution could explain the flat rotational
velocity curves(FRVC) of stars in galaxies because its energy density
is proportional to $1/r^2 $ and hence it can be dark matter in the
galatic halo.  Even if some questions are issued \cite{Que} about the
global monopole solution for FRVC, various generalized versions of the
global monopole were studied as models for dark matter and dark
energy \cite{Lee}.

Neutrino clustering was studied by some authors including one of
us \cite{BM} to explain the continuation of the cosmic ray spectrum
beyond the GZK(Greisen, Zatsepen and Kuzmin) cutoff. 
Neutrino-antineutrino($\nu$-$\overline{\nu}$) annihilation to a $Z^0
$-boson is one possible explanation for the phenomena.  But for
$\nu$-$\overline{\nu}$ annihilation into a $Z^0 $-boson to produce
super GZK air shower, a neutrino flux at $E_\nu \geq 10^{21} eV$ with
$m_\nu \sim$ a few $eV$ and a significant clustering of the relic
neutrino density in our galactic halo are required \cite{Wei}.

In this article we investigate if neutrinos can coalesce into neutrino
clouds in a curved spacetime.  Considering fermions coupled to an
$O(3)$ triplet of scalar fields in the most general static metric with
spherical symmetry, we show that only the total angular momentum
$j=1/2$ fermions can satisfy both Einstein's field equation and
Dirac's equation.  When the global $O(3)$ symmetry of the Lagrangian
is spontaneously broken to $U(1)$ at the ground state of the scalar
potential in the shape of Mexican hat, we can have a global monopole
solution similar to that of Barriola and Vilenkin.  Being coupled to a global
monopole situated at the galactic center, neutrinos are able to have
effective masses and  cluster in the galactic halo.

\section{Global $O(3)$ Symmetric Model for Neutrino Clustering}

The action of the global $O(3
)$ symmetric model of scalar fields
$\Phi^m $ and massive neutrinos $\Psi_n$ $(m, \, n=1,\, 2,\, 3)$
minimally coupled to gravity is given by
\begin{eqnarray}
%\begin{equation}
S&=&\int d^4 x \sqrt{-g}\,{\cal L} \, , \\
{\cal L}&=
& - \frac{1}{2} g^{\mu \nu}\partial_{\mu}\Phi^{m} \partial_{\nu}\Phi^{ m}
-V ({\vec{\Phi}}^2  ) \\ \nonumber
&+&\frac{i}{2}(\overline{\Psi}_n \gamma^a \nabla_a \Psi_n - \nabla_a \overline{\Psi}_n 
\gamma^a \Psi_n)-m_n \overline{\Psi}_n \Psi_n  \\ \nonumber
&+&\frac{g_y}{v} \Phi^m \Phi^m \overline{\Psi}_n \Psi_n         ,\nonumber
%\end{equation} 
\end{eqnarray}
where $V( {\vec{\Phi}}^2 )$ is a scalar potential and the last term is
the scalar-neutrino interaction for which we assume the first leading
order term preserving $O(3)$ symmetry, with the symmetry breaking
scale $v$ of the more fundamental theory.  Varying the action with
respect to the fields, we obtain following equations for scalar fields
$\Phi^{m}$ and neutrinos $\Psi_n$:
\begin{equation}
\frac{1}{\sqrt{-g}} \partial_{\mu}(\sqrt{-g} g^{\mu \nu}
\partial_{\nu} \Phi^m ) -\frac{\partial V }{\partial \Phi^{%
m}}+2\frac{g_y}{v}\overline{\Psi}_n \Psi_n \Phi^m =0\, ,
\end{equation}
\begin{equation}
i\gamma^a \nabla_a \Psi_n -m_n \Psi_n +\frac{g_y}{v}{\vec{\Phi}}^2
\Psi_n=0\, ,
\end{equation}
where the $\gamma^a$-matrices satisfy the Clifford algebra in a locally
flat inertial coordinate;
\begin{equation}
\{ \gamma^a ,\, \gamma^b  \}=-2\, \eta^{a b}
\end{equation}
with 
\begin{equation}
\eta^{a b}=Diag(-1,\,1,\,1,\,1) ,
\end{equation}
and the covariant derivative
\begin{equation}
\nabla_a ={e^\mu}_a (\partial_\mu +\Gamma_\mu )
\end{equation}
is constructed from the vierbein ${e^\mu}_a$ and spin connection
$\Gamma_\mu$, which we give explicitly in Eqs.  (25)-(31).

Using the standard definition of the energy-momentum tensor
\begin{equation}
T_{\mu \nu}\equiv -\frac{2}{\sqrt{-g}} \frac{\delta S}{\delta g^{\mu \nu}} \\
= -\frac{ e_{a \mu }}{ \det \{ e \} }\frac{\delta S}{\delta e_a^{\nu}}, 
\end{equation}
we have
\begin{eqnarray}
%\begin{equation}
T_{\mu \nu}&=&\partial_\mu \Phi^m \partial_\nu \Phi^m -g_{\mu
\nu}[\frac{1}{2}\partial^\beta \Phi^m \partial_\beta \Phi^m
+V({\vec{\Phi}}^2 ) ] \nonumber \\
&-&\frac{i}{4}[ (\overline{\Psi}_n
\gamma_\mu \nabla_\nu \psi_n -\nabla_\nu \overline{\Psi}_n \gamma_\mu
\Psi_n )+(\mu \leftrightarrow \nu)].
%\end{equation}
\end{eqnarray}
 The energy-momentum tensor allows us to construct the
Einstein equation,
\begin{equation}
G_{\mu\nu}=\kappa T_{\mu\nu},
\end{equation} 
where $G_{\mu \nu}$ is the Einstein tensor 
\begin{equation}
G_{\mu \nu}=R_{\mu \nu}-\frac{1}{2}g_{\mu \nu} R 
\end{equation}
with Ricci tensor $R_{\mu \nu}$ and $\kappa=8\pi G$.

When we consider the following potential of scalar fields
\begin{equation}
V({\vec{\Phi}}^2 )=\frac{\lambda}{4}({\vec{\Phi}}^2 -\eta^2  )^2 ,
\end{equation}
with a constant $ \eta$, the global $O(3)$ symmetry is spontaneously
broken to $U(1)$.  The scalar field configuration describing a global
monopole is known to be given by the hedgehog ansatz,
\begin{equation}
\Phi^m =F(r) \frac{x_m}{r},
\end{equation}
with a real function $F(r)$ and $r=(x_m x_m)^{1/2} =(x^2 +y^2 +z^2 )^{1/2}$,

Assume the line element of the spherically symmetric, static spacetime as
\begin{equation}
ds^2 =-\delta^2 (r) \alpha^2 (r) dt^2 +\frac{ dr^2}{ \alpha^{2} (r) }
+r^2 d\theta^2 +r^2 sin^2 \theta \, d\phi^2 .
\end{equation}
In the Cartesian coordinates,
\begin{equation}
x^{\mu}=(t, \, x^m )=(t,\,x,\,y,\,z), \,\,\,\,\, for\,\,\,\,  m=1,\,2,\,3,
\end{equation}
the line element can be written by
\begin{equation}
ds^2 =g_{\mu \nu}\, dx^\mu dx^\nu ,
\end{equation}
with the metric tensor
\begin{equation}
g_{\mu \nu}=-\delta^2 \alpha^2 \, {\delta^t}_\mu {\delta^t}_\nu
+[\delta_{m n}+(\frac{1}{ \alpha^{2}} -1)\frac{x_m x_n}{r^2}]
{\delta^m}_\mu {\delta^n}_\nu
\end{equation}
and its inverse
\begin{equation}
g^{\mu \nu}=- \frac{1}{\delta^{2} \alpha^{2}} {\delta^\mu}_t {\delta^\nu}_t +[\delta_{m n}+(\alpha^2 -1)\frac{x_m x_n}{r^2}]
 {\delta^\mu}_m {\delta^\nu}_n .
\end{equation}

From the standard definition of Christoffel symbols
\begin{equation}
{\Gamma}_{\alpha \beta}^\mu =\frac{1}{2}g^{\mu
\nu}(g_{\nu\alpha,\,\beta}+g_{\nu\beta,\,\alpha} -g_{\alpha
\beta,\,\nu} ),
\end{equation}
we derive
\begin{eqnarray}
&{\Gamma}_{t \, m}^t &=\frac{x_m}{2 r }(\frac{{\delta^2
,}_{\,r}}{\delta^2} +\frac{{\alpha^2 ,}_{\,r}}{\alpha^2 }), \nonumber 
\\
&{\Gamma}_{t \, t}^l &=\frac{x_l \, \alpha^2}{2 r }(\delta^2 {\alpha^2
,}_{\,r}+ {\delta^2 ,}_{\,r}\, \alpha^2 ), \\ \nonumber &{\Gamma}_{m
n}^l &=\frac{x_l \, \alpha^2 }{2 r }\bigl[ -\frac{{\alpha^2
,}_{\,r}}{\alpha^4}\frac{x_m x_n}{r^2} +(\alpha^{-2} -1)(\delta_{m
n}-\frac{x_m x_n}{r^2})\frac{2}{r} \bigr] , \\
&\mbox{Others}&=0.\nonumber
\end{eqnarray}
In the metric given in Eq. (14), scalar field equation reads
\begin{eqnarray}
&&\alpha^2 {F,}_{\, rr}+\frac{\alpha^2 }{2}{F,}_{\, r}(\frac{{\delta^2
,}_{\,r}}{\delta^2} +2\frac{{\alpha^2 ,}_{\,r}}{\alpha^2 }
+\frac{4}{r}) -\frac{2F}{r^2 } \nonumber \\
&&-\frac{\partial V}{\partial F}+\frac{2g_y}{v} \overline{\Psi}_n \Psi_n F= 0\, ,
\end{eqnarray}
where ${F,}_{\, r}\equiv \frac{\partial F}{\partial r},\, {F,}_{\, rr}\equiv \frac{\partial^2 F}{\partial r^2 }, ...$ .

We rewrite the line element as

\begin{equation}
ds^2 =\eta_{a b}\, e^a e^b
\end{equation}
with inverse of $\eta^{a b}$,
%\begin{equation}
$\eta_{a b} =Diag(-1,\,1 ,\,1,\,1) $
%\end{equation}
and 
\begin{equation}
e^0 =\alpha \delta \, dt, \, e^1 =\frac{dr}{\alpha},\, e^2 =r \, d\theta,\, 
e^3 =r\, sin\theta\, d\phi .
\end{equation}
Putting
\begin{equation}
e^a ={e^a}_{\mu}\, dx^{\mu} ,
\end{equation}
we find the vierbein given by
\begin{eqnarray}
{e^{a}}_{\mu} =
\left(
\begin{array}{cccc}
\alpha \delta  & 0 & 0 & 0 \\ 
0 & \frac{1}{\alpha} sin \theta cos \phi & \frac{1}{\alpha} sin \theta
sin \phi & \frac{1}{\alpha} cos \theta \\
0 & cos \theta cos \phi & cos \theta sin\phi & -sin\theta \\
0 & -sin \phi & cos\phi & 0 
\end{array}
\right)
\end{eqnarray}
and their inverses
\begin{eqnarray}
{e^{\nu}}_{b}& =&\left(
\begin{array}{cccc}
\frac{1}{ \alpha \delta} & 0 & 0 & 0 \\
0 & \alpha sin \theta cos \phi & cos \theta cos \phi & -sin\phi \\
0 & \alpha sin \theta sin \phi & cos \theta sin\phi & cos\phi \\
0 & \alpha cos \theta & -sin\theta & 0
\end{array}
\right)\\&=&
\left(
\begin{array}{cccc}
\frac{1}{\alpha \delta  }&0&0&0\\
0&(\alpha\frac{\partial x^m}{\partial r}) &(\frac{1}{r}\frac{\partial
x^m}{\partial \theta}) &(\frac{1}{r\, sin\theta}\frac{\partial
x^m}{\partial \phi})
\end{array}
\right).
\end{eqnarray}
The vierbein satisfy the relations 
\begin{equation}
{e^a}_{\mu} {e^{\mu}}_b ={{\delta}^a}_b ,
\end{equation}
\begin{equation}
 {e^{\nu}}_a {e^a}_{\mu}={\delta^{\nu}}_{\mu} .
\end{equation}
The spin connection which appeared in Eq. (7) is defined as
\begin{equation}
 \Gamma_\mu =-\frac{1}{4}\gamma^a \gamma^b \, e_{\nu a} \, {e^\nu}_{b;\,  \mu}
\end{equation}
with
\begin{equation}
{e^\nu}_{b;\, \alpha} = {e^\nu}_{b,\, \alpha}+{\Gamma}_{\alpha
\beta}^\nu \, {e^\beta}_b .
\end{equation}
We calculate the spin connections from the above  Eq. (20)
and Eq. (30), and obtain
\begin{eqnarray}
\Gamma_t & =&\frac{1}{4}\gamma^0 \gamma^1 ({2\alpha^2 {\delta,}_{\, r}
+\delta {\alpha^2},}_{\, r}) \\ \nonumber \Gamma_x
&=&\frac{1}{2}\gamma^1 \gamma^2 \frac{\alpha \, cos \theta cos
\phi}{r} -\frac{1}{2}\gamma^2 \gamma^3 \frac{cos\theta sin \phi}{r
sin\theta} \nonumber \\
&+&\frac{1}{2}\gamma^3 \gamma^1 \frac{\alpha \, sin\phi}{r}, \nonumber
\\
\Gamma_y &=&\frac{1}{2}\gamma^1 \gamma^2 \frac{\alpha \, cos \theta
sin \phi}{r} +\frac{1}{2}\gamma^2 \gamma^3 \frac{cos\theta cos \phi}{r
sin\theta} \nonumber \\
&-&\frac{1}{2}\gamma^3 \gamma^1 \frac{\alpha \, cos\phi}{r},\nonumber \\
\nonumber \Gamma_z &=&-\frac{1}{2}\gamma^1 \gamma^2 \frac{\alpha \,
sin\theta}{r} ,\nonumber
\end{eqnarray}
which satisfy the Cartan structure equation,
\begin{equation}
[\gamma^\nu , \,\Gamma_\mu ]={\gamma^\nu }_{;\, \mu} .
\end{equation}

\section{Spherical Symmetry and $j=\frac{1}{2}$ Fermions}

We can show that neutrino field equations given by 
\begin{equation}
i\gamma^a {e^{\mu}}_a (\partial_\mu +\Gamma_\mu ) \Psi_n -m_n \Psi_n
+\frac{g_y}{v}{\vec{\Phi}}^2 \Psi_n =0
\end{equation}
 become
\begin{eqnarray}
&&\gamma^0 \frac{i}{\alpha\delta } \partial_t \Psi_n + \gamma^1 i
\alpha \hat{D}_r \Psi_n
+\gamma^2 \frac{i}{r} (
\partial_{ \theta}+\frac{cos\theta}{2 sin\theta})\Psi_n \nonumber \\
&&+\gamma^3
\frac{i}{r\, sin\theta} \partial_{\phi}\Psi_n -(m_n -
\frac{g_y}{v}{\vec{\Phi}}^2 )\Psi_n=0 \, ,
\end{eqnarray}
in the spherically symmetric and static metric in Eq. (14).
Equivalently, we can write the last equation as
\begin{equation}
i\partial_t \Psi_n =\hat{H} \Psi_n \, ,
\end{equation}
where the Hamiltonian is defined as
\begin{equation}
\hat{H}=\alpha\delta \bigl[ \gamma^0 \gamma^1 \frac{ \alpha}{i}
\hat{D}_r +\gamma^1 \frac{1}{ir} \hat{k} +\gamma^0 (m_n -
\frac{g_y}{v}{\vec{\Phi}}^2 )\bigr] ,
\end{equation}
with
\begin{equation}
\hat{D}_r =
{\partial_r}+\frac{1}{r}+\frac{1}{4}(\frac{{{\delta^2},}_{\,
r}}{\delta^2}+\frac{{{\alpha^2},}_{\, r}}{\alpha^2} ).
\end{equation}
Here we have defined the operator $\hat{k}$
as\cite{Dirac} \cite{Bou}
\begin{equation}
\hat{k}\equiv i\gamma^0 \gamma^1 \bigl( \gamma^2 \frac{1}{i \sqrt{sin
\theta}} \partial_{\theta} \sqrt{sin \theta} +\gamma^3 \frac{1}{i\,
sin\theta} \partial_{\phi} \bigr),
\end{equation}
which commutes with the Hamiltonian operator $\hat{H}$
\begin{equation}
[\hat{k},\,\hat{H}]=0. 
\end{equation}

Taking the representation
 of $\gamma$-matrices to be direct products 
of independnet Pauli matrices $\sigma_{i}, \rho_{j}$($\vec{\rho}\otimes \vec{\sigma}$ representation) \cite{Dirac} \cite{Bou},
\begin{equation}
\gamma^0 =\rho_2 ,\, \gamma^1 =i \rho_1 ,\, \gamma^2 =-i \rho_3
\sigma_3 , \, \gamma^3 =-i \rho_3 \sigma_1 ,
\end{equation}
 we can represent $\hat{k}$ in 2-dimensional $\sigma$-space
\begin{equation}
\hat{k}= 
\sigma_3 \frac{1}{i \sqrt{ sin \theta}} \partial_{\theta} \sqrt{sin \theta}
+\sigma_1 \frac{1}{i \,sin\theta} \partial_{\phi} \, ,
\end{equation}
and we can solve the eigenvalue equation,
\begin{equation}
\hat{k} {\cal Y}_k^m (\theta, \phi)={k} {\cal Y}_k^m (\theta, \phi) \, ,
\end{equation}
with 2-dimensional
spinor spherical harmonics \cite{Bou}
\begin{eqnarray}
&&{\cal Y}_k^m (\theta, \phi)=\frac{e^{im\phi}}{ \sqrt{2 \pi}}\biggl[
\frac{(j+m)!}{(j-m)!} { \biggr]}^{1/2}
\frac{(tan\frac{\theta}{2})^{{\sigma_2}/2}}{sin^m \theta} \times \nonumber \\
&&\biggl( \frac{\partial}{\partial cos\theta}{\biggr)}^{j-m}\frac{
sin^{2j}\theta }{ 2^j (j-\frac{1}{2})!  } \bigl(
tan\frac{\theta}{2}{\bigr)}^{-\sigma_2} u_k ,
\end{eqnarray}
where
\begin{eqnarray}
u_k =\frac{1}{\sqrt{2} \arrowvert k\arrowvert }\left(
\begin{array}{c}
\arrowvert k\arrowvert \\k
\end{array}
\right)
\end{eqnarray}
with $k=\pm (j+\frac{1}{2})$ and $j=l \pm \frac{1}{2}$.  In 
the representation (41) for $\gamma$-matrices,  the Hamiltonian is
expressed in terms of 2-dimensional $\rho$-matrices
\begin{equation}
\hat{H}= \alpha \delta\bigl[ \rho_3  \frac{ \alpha}{i} \hat{D}_r  
+\rho_1 \frac{1}{r} \hat{k} +\rho_2  (m_n - \frac{g_y}{v}{\vec{\Phi}}^2 ) \bigr]
\end{equation}

 The eigenfunctions of both $\hat{H}$ and $\hat{k}$ can be written as
 direct products of the $\rho$-space spinor ${\psi_n^{(k m)}}$ and
 $\sigma$-space spinor ${\cal Y}_k^m $,
\begin{equation}
\Psi_n (t,r,\theta,\phi)=e^{-i E_n t} \sum_{-j\le m\le j} {\psi_n^{(k
m)}} (r)\, \, {\cal Y}_k^m (\theta, \phi) ,
\end{equation}
and the neutrino equations $i\partial_t \Psi_n =\hat{H} \Psi_n$ then read
\begin{equation}
\frac{E_n}{\alpha\delta }{\psi_n^{(km)}}=\rho_3 \frac{ \alpha}{i}
\hat{D}_r {\psi_n^{(km)}} +\bigl[ \rho_1 \frac{k}{r} +\rho_2 (m_n -
\frac{g_y}{v}{\vec{\Phi}}^2 )\bigr]\, {\psi_n^{(km)}} .
\end{equation}
Moreover, if we put
\begin{equation}
{\psi_n^{(km)}} (r) \equiv \frac{1}{r}(\alpha^2 \delta^2 )^
\frac{-1}{4} \,  
{R_n^{(km)}} (r),
\end{equation}
since
\begin{equation}
\hat{D}_r {\psi_n^{(km)}} = \frac{1}{r}(\alpha^2 \delta^2 )^
\frac{-1}{4}\, \partial_r {R_n^{(km)}} , 
\end{equation}
then we get
\begin{equation}
\frac{E_n}{\alpha \delta }{R_n^{(km)}}=\rho_3 \frac{ \alpha}{i}
{\partial}_r {R_n^{(km)}} +[\rho_1 \frac{k}{r} +\rho_2 (m_n -
\frac{g_y}{v}{\vec{\Phi}}^2 )]{R_n^{(km)}} .
\end{equation}
Setting 
\begin{eqnarray}
{R_n^{(km)}} =\left(
\begin{array}{c}
{\varphi_n^{(km)}} \\
{\chi_n^{(km)}}
\end{array}
\right),
\end{eqnarray}
we obtain the  coupled equations,
\begin{eqnarray}
(\frac{E_n}{\alpha \delta} +i \alpha\partial_r )\,{\varphi_n^{(km)}}  & =&[
\frac{k}{r} -i(m_n -\frac{g_y}{v} {\vec{\Phi}}^2 ) ]\, {\chi_n^{(km)}} \\
(\frac{E_n}{\alpha \delta} -i \alpha\partial_r ){\chi_n^{(km)}} &
=&[\frac{k}{r} +i(m_n -\frac{g_y}{v} {\vec{\Phi}}^2 ]\,
{\varphi_n^{(km)}},  \nonumber
\end{eqnarray}
asymptotic solutions to which are given in the following section.

Einstein's equation 
$G_{\mu\nu}=\kappa T_{\mu\nu}$
can be rewriten as
\begin{equation}
R_{\mu\nu}=\kappa(T_{\mu\nu}-\frac{1}{2}g_{\mu\nu}  {T_{\alpha}}^{\alpha}) .
\end{equation}
Substituting Eq. (9) into the Einstein's equation in Eq. (54) 
gives
\begin{eqnarray}
R_{\mu\nu}=&& \kappa [\partial_{\mu}\Phi^{m}\partial_{\nu}\Phi^{ m}+
g_{\mu\nu} V +\frac{i}{4} g_{\mu \nu} \Sigma\\ \nonumber
&&-\frac{i}{4}(\overline{\Psi}_n \gamma_\mu \nabla_\nu \Psi_n
-\nabla_\mu \overline{\Psi}_n \gamma_\nu \Psi_n \\ \nonumber
&&+\overline{\Psi}_n
\gamma_\nu \nabla_\mu \Psi_n -\nabla_\nu \overline{\Psi}_n \gamma_\mu
\Psi_n)],
\end{eqnarray}
with
\begin{equation}
\Sigma={e^\alpha}_c (\overline{\Psi}_n \gamma^c \partial_\alpha \Psi_n -
\partial_\alpha \overline{\Psi}_n \gamma^c  \Psi_n ) .
\end{equation}
Elements of Ricci tensor calculated from metric coefficients in
Eq.  (14) and the ansatz for $\Phi^m $ in Eq.  (13) are substituted
into Eq.  (55), to give the set of equations
\begin{eqnarray}
&&\frac{1}{2 \delta^2 }{(\delta^2 \alpha^2 ) ,}_{\, r r }
-\frac{{(\delta^2 \alpha^2 ),}_{\, r}} {4\delta^2 }\frac{{\delta^2
,}_{\, r}}{\delta^2 } + \frac{1}{r} \frac{{(\delta^2 \alpha^2 ) ,
}_{\, r }}{\delta^2 }= \nonumber \\
&&\kappa[ -V -\frac{i}{4} \Sigma +
\frac{i}{2\delta \alpha}(\overline{\Psi}_n \gamma^0 \partial_t \Psi_n
-\partial_t \overline{\Psi}_n \gamma^0 \Psi_n) ],
\end{eqnarray}

\begin{eqnarray}
&&-\frac{1}{2 \delta^2 }{(\delta^2 \alpha^2 ) ,}_{\, r r }
+\frac{{(\delta^2 \alpha^2 ),}_{\, r} }{4\delta^2 }\frac{{\delta^2
,}_{\, r}}{\delta^2 } - \frac{1}{r}{\alpha^2 , }_{\, r }= \nonumber \\
&&\kappa[\alpha^2 ({F,}_{\, r})^2 + V +\frac{i}{4} \Sigma \nonumber\\
&& -\frac{i
\alpha}{2} (\overline{\Psi}_n \gamma^1 \partial_r \Psi_n -\partial_r
\overline{\Psi}_n \gamma^1 \Psi_n)],
\end{eqnarray}

\begin{eqnarray}
&& \frac{1}{r^2}{(1-\alpha^2 ) }- \frac{1}{r}{\alpha^2 ,}_{\, r }-
 \frac{\alpha^2 }{2r}\frac{{\delta^2 , }_{\, r }}{\delta^2 }= \nonumber\\
&& \kappa[\frac {F^2}{r^2} + V +\frac{i}{4} \Sigma \nonumber \\
&& - \frac{
 i}{2r}(\overline{\Psi}_n \gamma^2 \partial_\theta \Psi_n
 -\partial_\theta \overline{\Psi}_n \gamma^2 \Psi_n) ] ,
\end{eqnarray}

\begin{eqnarray}
&& \frac{1}{r^2}{(1-\alpha^2 ) }- \frac{1}{r}{\alpha^2 ,}_{\, r }-
 \frac{\alpha^2 }{2r}\frac{{\delta^2 , }_{\, r }}{\delta^2 }= \nonumber \\
&& \kappa[\frac {F^2}{r^2} + V +\frac{i}{4} \Sigma \nonumber \\
&&- \frac{ i}{2 r\,
 sin\theta}(\overline{\Psi}_n \gamma^3 \partial_\phi \Psi_n
 -\partial_\phi \overline{\Psi}_n \gamma^3 \Psi_n) ] .
\end{eqnarray}
 The first in above equations is the ($t, t$)-component of Eq.  (55),
 the second ($r, r$)-component, the third ($\theta,
 \theta$)-component, and the last ($\phi, \phi$)-component,
 respectively.

($\theta, \theta$)-component of the Einstein equation is same as
($\phi, \phi$)-component of Einstein equation if
\begin{equation}
\overline{\Psi}_n \gamma^2 \partial_\theta \Psi_n
-\partial_\theta \overline{\Psi}_n \gamma^2  \Psi_n =
\frac{1}{sin\theta }(\overline{\Psi}_n \gamma^3 \partial_\phi \Psi_n
-\partial_\phi \overline{\Psi}_n \gamma^3  \Psi_n ).
\end{equation}
 Using $\vec{\rho}\otimes\vec{\sigma}$ representation for $\gamma$-matrices and the
 mathematical identities;
\begin{eqnarray}
\sigma_3 \partial_\theta {\cal Y}_k^m =&&(\pm \sigma_3 m\, cot\theta \pm \sigma_1\frac{i}{2sin\theta} ){\cal Y}_k^m \nonumber \\
&&\pm \sigma_3 e^{\mp i \phi}\sqrt{(j\pm m+1)(j\mp m)}{\cal Y}_k^{m\pm 1} , 
\end{eqnarray}
this  condition becomes
\begin{eqnarray}
&&\sum_{m,m'}\psi_n^{\dagger (km')} \rho_1 \psi_n^{(km)} \times  \nonumber \\
&& \biggl[ {\cal Y}_k^{\dagger m'}\{ \pm \sigma_3 (m-m') cot\theta \pm
\sigma_1\frac{i}{sin\theta} \} {\cal Y}_k^m \\ \nonumber 
&&\pm {\cal
Y}_k^{\dagger m'}\sigma_3 e^{\mp i \phi}\sqrt{(j\pm m+1)(j\mp m)}{\cal
Y}_k^{m\pm 1} \\ \nonumber 
&&\mp {\cal Y}_k^{\dagger m'\pm 1}\sigma_3
e^{\pm i \phi}\sqrt{(j\pm m'+1)(j\mp m')}{\cal Y}_k^{m} \biggr]\\
\nonumber 
&&=\sum_{m,m'}\psi_n^{\dagger (km')} \rho_1
\psi_n^{(km)}(m+m'){\cal Y}_k^{\dagger m'} \sigma_1\frac{i}{sin\theta}
{\cal Y}_k^m  .
\end{eqnarray}
The condition (63) can be satisfied if
\begin{eqnarray}
m&=&m' ,\\ \nonumber
m+m' & =&2m=\pm 1 ,\\ \nonumber
j&=&\frac{1}{2} ,
\end{eqnarray}
since
\begin{equation}
{\cal Y}_{k(j=\frac{1}{2})}^{\pm\frac{1}{2} \pm 1 } =0 .
\end{equation}
Moreover, explicit calculations show us that
\begin{eqnarray}
\bigl( {\cal Y}_{k(j=\frac{1}{2})}^{ \pm\frac{1}{2}}\bigr)^{\dagger}
\,\, {\cal Y}_{k(j=\frac{1}{2})}^{\pm\frac{1}{2}}&=&\frac{1}{4\pi},\\
\nonumber \bigl( {\cal Y}_{k(j=\frac{1}{2})}^{ \pm
\frac{1}{2}}\bigr)^{\dagger} \,\, \sigma_1 \,\, {\cal
Y}_{k(j=\frac{1}{2})}^{\pm\frac{1}{2}} &=& \pm
\frac{sin\theta}{4\pi}\frac{k}{\arrowvert k\arrowvert} .
\end{eqnarray}
These make Eqs.  (57)-(60) $\theta$- and $\phi$-independent, which is
consistent with spherically symmetric metric in Eq.  (14), when we
consider only ($j=\frac{1}{2},\, m=\frac{1}{2}$) or ($j=\frac{1}{2},
\,m=-\frac{1}{2}$) fermions.

For the case ($j=\frac{1}{2},\, m=\frac{1}{2}$) or ($j=\frac{1}{2},
\,m=-\frac{1}{2}$) with $k=\pm 1 $, neutrino wave function is
\begin{equation}
\Psi_n (t,r,\theta,\phi)=e^{-i E_n t} {\psi_n^{(k,\, \pm\frac{1}{2})}}
(r)\, \, {\cal Y}_k^{\pm \frac{1}{2}} (\theta, \phi) .
\end{equation}
Let $\psi_n (r) \equiv {\psi_n^{(k, \,\frac{1}{2})}} (r)$ or $
{\psi_n^{(k,\, -\frac{1}{2})}} (r)$ and
\begin{eqnarray}
{\psi_n} (r) =\frac{1}{r}(\alpha^2 \delta^2 )^{-\frac{1}{4}}\left(
\begin{array}{c}
{\varphi_n} \\
{\chi_n}
\end{array}
\right).
\end{eqnarray}
In this case the term $\Sigma$ in Eq. (56) reads 
\begin{equation}
\Sigma =-\frac{i}{r^2 \alpha^2 \delta^2}q_0 +\frac{i}{r^2 \delta}q_1 +
\frac{i}{r^2 \alpha \delta} q_k ,
\end{equation}
with
\begin{equation}
q_0 =\frac{2E_n}{4\pi}r^2 \alpha\delta \, \psi_n^\dagger \psi_n
=\frac{2E_n}{4\pi}(\varphi_n^* \varphi_n +\chi_n^* \chi_n) ,
\end{equation}
\begin{eqnarray}
 q_1 && =\frac{-i}{4\pi}r^2 \alpha\delta (\psi_n^\dagger \rho_3 \partial_r
\psi_n -\partial_r \psi_n^\dagger \rho_3 \psi_n ) \nonumber \\
&& =\frac{-i}{4\pi}
(\varphi_n^* \partial_r \varphi_n -\partial_r \varphi_n^* \varphi_n \\
&& -\chi_n^* \ \partial_r \chi_n+\partial_r \chi_n^* \chi_n) ,\nonumber
\end{eqnarray}
\begin{equation}
q_k =\frac{1}{2\pi}r \alpha\delta \frac{k}{|k|}\psi_n^\dagger \rho_1
\psi_n =\frac{1}{2\pi r}\frac{k}{|k|}(\chi_n^* \varphi+\varphi_n^*
\chi_n) ,
\end{equation}
\begin{equation}
q_2 =\frac{1}{4\pi}r^2 \alpha\delta \,\psi_n^\dagger \rho_2 \psi_n
=\frac{i}{4\pi}(\chi_n^* \varphi-\varphi_n^* \chi_n) ,
\end{equation}
Using the last equations  Einstein's equations are given by
\begin{eqnarray}
&&\frac{1}{2 \delta^2 }{(\delta^2 \alpha^2 ) ,}_{\, r r }
-\frac{{(\delta^2 \alpha^2 ),}_{\, r}} {4\delta^2 }\frac{{\delta^2
,}_{\, r}}{\delta^2 } + \frac{1}{r} \frac{{(\delta^2 \alpha^2 ) ,
}_{\, r }}{\delta^2 } \nonumber \\ 
&&=\kappa[ -V+ \frac{1}{4 r^2}(\frac{q_0}{\delta^2
\alpha^2}+\frac{q_1}{\delta}+\frac{q_k}{\alpha\delta}) ],
\end{eqnarray}

\begin{equation}
\frac{\alpha^2 }{r }\frac{{\delta^2 ,}_{\, r}}{\delta^2 }
=\kappa[\alpha^2 ({F,}_{\, r})^2 +\frac{1}{2r^2}(\frac{q_0}{\delta^2
\alpha^2}+\frac{q_1}{\delta} )],
\end{equation}

\begin{equation}
 \frac{1}{r^2}{(1-\alpha^2 ) }- \frac{1}{r}{\alpha^2 ,}_{\, r }-
 \frac{\alpha^2 }{2r}\frac{{\delta^2 , }_{\, r }}{\delta^2 }
 =\kappa[\frac {F^2}{r^2} + V +\frac{1}{4r^2}(\frac{q_0}{\delta^2
 \alpha^2}-\frac{q_1}{\delta} ) ] ,
\end{equation}
and scalar field equation 
\begin{eqnarray}
&&\alpha^2 {F,}_{\, rr}+\frac{\alpha^2 }{2}{F,}_{\, r}(\frac{{\delta^2
,}_{\,r}}{\delta^2} +2\frac{{\alpha^2 ,}_{\,r}}{\alpha^2 }
+\frac{4}{r})-\frac{2F}{r^2 } \nonumber \\
&&-\frac{\partial V}{\partial F}+\frac{2g_y}{v} \frac{q_2}{r^2\delta\alpha} F =0\, .
\end{eqnarray}

\section{Asymptotic Solutions for Large $r$}

As the global monopole solution \cite{Vil}, one component $\alpha (r)$
of the metric in Eq.  (14), far away from the galactic core, can be
asymptotically taken as
\begin{equation}
\alpha^2 \simeq {{\alpha}_o}^2 \equiv c -\frac{2M}{r},
\end{equation}
with a constant $c$ when
\begin{equation}
F\simeq \eta, \,\,\,\,\, V(\eta^2)\simeq 0,\,\,\,\,\, \frac{\partial
V}{\partial F} (\eta^2)\simeq 0.
\end{equation}
Analysing Eqs.  (74)-(77) with the assumption that $q_i $'s in Eqs. 
(70)-(73) are ${\cal O}(r^0)$, we can get an asymptotic solution for
another component of the metric in Eq.  (14) as
\begin{equation}
\delta^2 \simeq A^2 ln(\frac{r}{2M}),
\end{equation}
with a constant $A$.
In this limit with the help of Eq. (68) the neutrino equation (53) becomes
\begin{equation}
{\partial_r}^2 \varphi_n \simeq -q^2 (r) \varphi_n ,
\end{equation}
with
\begin{equation}
q(r) \equiv \sqrt{ \frac{ {E_n}^2 }{ \delta^2 \,{\alpha_o}^4 }-
\frac{ {{m^2}_{o.\, n} } } {{\alpha_o}^2}  }
\end{equation}
and ${m}_{o.\,n}={m}_n -\frac{g_y}{v}\eta^2 $.

In the region
\begin{equation}
0<ln(\frac{r}{2M})\leq \frac{{E_n}^2 }{A^2 {\alpha_o}^2  {{m^2}_{o.\, n}} }
\end{equation}
we have an asymptotic solution to Eq. (81),
\begin{eqnarray}
\varphi_n & \simeq & A_n e^{i q(r) r} + B_n e^{-i q(r) r} , \nonumber \\
 \chi_n &\simeq & \frac{ i}{m_{o.\, n}}[A_n (\frac{ {E_n} }{
 {\alpha_o} \delta }-\alpha_o q(r)) e^{+i q(r) r} \\
&+& B_n (\frac{ {E_n}
 }{ {\alpha_o} \delta }+\alpha_o q(r)) e^{-i q(r) r}]. \nonumber
\end{eqnarray}
With above asymptotic solutions to Dirac's equation for neutrinos, the
$q_i $ in Eqs.  (70)-(73) are given by
\begin{eqnarray}
q_0 &\simeq& \frac{E_n }{\pi}[{B^2}_n (1+\frac{1}{{m^{2}}_{o.\,
n}}(\frac{ {E_n}^2 }{ {\alpha_o}^2 {\delta}^2 }+{\alpha_o}^2 q^2 (r))
) \nonumber \\
& +&A_n B_n (1+ {m^2}_{o.\, n}) cos(2q r)], \nonumber \\
q_1 &\simeq&\frac{2E_n }{\pi}{B^2}_n \frac{q^2}{\delta} ,\nonumber \\
q_k &\simeq&\frac{2 k}{\pi r|k|}A_n B_n \frac{\alpha_o q}{m_{o.\, n}}
sin(2q r), \nonumber\\
q_2 &\simeq&\frac{E_n }{\pi m_{o.\, n} \alpha_o \delta } [ A_n B_n
cos(2q r) +{B_n}^2 ], \nonumber
\end{eqnarray}
with real constants $A_n$ and $B_n $ such that $|A_n| =|B_n|$.  The
$q_i$'s are ${\cal O}(r^0)$ or less and so are  consistent with the
assumption of  Eq.  (80).  For $r>r_o$ with
$ln(\frac{r_o}{2M})\equiv \frac{{E_n}^2 }{A^2 {\alpha_o}^2
{{m^2}_{o.\, n}} }$ , $q(r)$ in Eq.  (82) is pure imaginary and then
wave functions of neutrinos become to be multiplied by exponentially
decaying factors.  Therefore we get {\it neutrino clustering} with the
radius $r_o$.  Taking $r_o \simeq 10^{23} cm$ which corresponds to the
radius of our Galaxy and assuming that there exists a supermassive
black hole at the center of Galaxy with a mass $M_{SBH}\simeq 3\times
10^6 M_{\odot}$ \cite{BH}, we can estimate the energy of neutrinos at
$E_n \simeq 5\,m_{o.  n}$ , where we have used the relations, $
\frac{{E_n}^2 }{A^2 {\alpha_o}^2 {{m^2}_{o.\, n}} }\simeq
\ln (\frac{r_o}{2M_{SBH}}) $ and $A^2 {\alpha_o}^2 \simeq {\cal O}(1 )$.

\section{Asymptotic Solutions for Small $r$}
Next let us study the small $r $ behaviors of components of the metric
in Eq.  (14), $\alpha$ and $\delta$.  For small $r$ near the Galactic
core we adopt the Thomas-Fermi(TF) approximation for the energy
density and the pressure of neutrinos, as in the case of fermion
stars \cite{TDL}\cite{Q}\cite{QGM}. From the local conservation law,
$0=(\overline{\Psi}\gamma^{\mu}\Psi)_{;\,
\mu}=\frac{1}{\sqrt{-g}}\partial_{\mu}(\sqrt{-g}\,
\overline{\Psi}\gamma^{\mu}\Psi )$, we have the conserved, total
number of neutrinos,
\begin{equation}
N=\int d^3 x\sqrt{-g}\, \overline{\Psi}\gamma^{t}\Psi = \int d^3
x\sqrt{-g}{\Psi}^ {\dagger}\Psi \,{e^t}_0 .
\end{equation}
 The number density of neutrinos in the TF approximation is given by
\begin{equation}
\big< \Psi^{\dagger} \Psi \big>_{TF} =\frac{2}{(2\pi)^3 }\int d^3 q\,
n_q =\frac{{q_F}^3 (r)}{3\pi^2}\,,
\end{equation}
where the Fermi distribution $n_q =\theta (q_F -q)$ with the fermi
momentum $q_F$. From Eq.  (86) and the time-independent
Dirac's equation, $i\partial_t \Psi_n=E_n \Psi_n$, we can put
\begin{eqnarray}
&& \big< \frac{i}{2\delta \alpha}(\overline{\Psi}_n \gamma^0 \partial_t
 \Psi_n -\partial_t \overline{\Psi}_n \gamma^0 \Psi_n ) {\big>}_{TF} \nonumber \\
&& =\frac{2}{(2\pi)^3 }\int d^3 q\, n_q\, E_n (q) \,{e^t}_0 \equiv \rho
 (r) ,
\end{eqnarray}
where $\rho$ is the energy density of neutrinos and ${e^t}_0 =
\frac{1}{\delta \alpha}$.  
%We define the radial pressure of neutrinos
%as
The stress tensor of neutrinos has the following diagonal elements in the vierbein basis: 
\begin{eqnarray}
\big< \frac{ \alpha}{2 i} (\overline{\Psi}_n \gamma^1 \partial_r \Psi_n
-\partial_r \overline{\Psi}_n \gamma^1  \Psi_n ) {\big>}_{TF} & =&p_1 , \\
\big<  \frac{ 1}{2 i r}(\overline{\Psi}_n \gamma^2 \partial_\theta
\Psi_n -\partial_\theta \overline{\Psi}_n \gamma^2 \Psi_n){\big>}_{TF}
&=&p_2 , \nonumber 
\end{eqnarray}
and
\begin{equation}
 \big< \frac{ 1}{2 i r\, sin\theta}(\overline{\Psi}_n \gamma^3
 \partial_\phi \Psi_n -\partial_\phi \overline{\Psi}_n \gamma^3 \Psi_n
 ){\big>}_{TF} = p_{3} ,
\end{equation}
respectively.

We assume, in the spherically symmetric spacetime we consider, that
$p_1 =p_2 =p_3 \equiv p(r)$.
From Dirac's equation (36) and its Hermitian conjugate, we have the relation;
\begin{eqnarray}
&&\frac{i}{\delta \alpha}(\overline{\Psi}_n \gamma^0 \partial_t \Psi_n
-\partial_t \overline{\Psi}_n \gamma^0 \Psi_n ) \nonumber \\
&& +{i \alpha}
(\overline{\Psi}_n \gamma^1 \partial_r \Psi_n -\partial_r
\overline{\Psi}_n \gamma^1 \Psi_n ) \nonumber \\
&&+ \frac{ i}{r}(\overline{\Psi}_n
\gamma^2 \partial_\theta \Psi_n -\partial_\theta \overline{\Psi}_n
\gamma^2 \Psi_n)  \\
&&+ \frac{ i}{ r\, sin\theta}(\overline{\Psi}_n \gamma^3 \partial_\phi \Psi_n
-\partial_\phi \overline{\Psi}_n \gamma^3  \Psi_n )
= 2 m_{eff} (r) \,\overline{\Psi}_n \Psi_n , \nonumber
\end{eqnarray}
with $m_{\rm {eff}} (r)=m_{n}-\frac{g_y}{v} F^2 (r) $.
Since the left hand side of above equation is same as $i \Sigma$ in Eq. (56),
we have the following relation in the TF approximation.
\begin{equation}
\frac{i}{2}\big< \Sigma{\big>}_{TF} \equiv \rho -3 p = m_{\rm{eff}}\,
\frac{v}{2 g_y} Q ,
\end{equation}
where we have defined
$  \frac{v}{2 g_y} Q (r) \equiv \big< \overline{\Psi}_n \Psi_n
{\big>}_{TF}   $. 
Thus in the TF approximation the Einstein equations (57)-(60) read

\begin{eqnarray}
&&\frac{1}{2 \delta^2 }{(\delta^2 \alpha^2 ) ,}_{\, r r }
-\frac{{(\delta^2 \alpha^2 ),}_{\, r}} {4\delta^2 }\frac{{\delta^2
,}_{\, r}}{\delta^2 } + \frac{1}{r} \frac{{(\delta^2 \alpha^2 ) ,
}_{\, r }}{\delta^2 } \nonumber \\ &&=\kappa[ -V+\frac{1}{2}\rho +\frac{3}{2}p ],
\end{eqnarray}

\begin{eqnarray}
&&-\frac{1}{2 \delta^2 }{(\delta^2 \alpha^2 ) ,}_{\, r r }
+\frac{{(\delta^2 \alpha^2 ),}_{\, r} }{4\delta^2 }\frac{{\delta^2
,}_{\, r}}{\delta^2 } - \frac{1}{r}{\alpha^2 , }_{\, r }
\nonumber \\& &=\kappa[\alpha^2 ({F,}_{\, r})^2 + V +\frac{1}{2}\rho -\frac{1}{2}p ],
\end{eqnarray}

\begin{eqnarray}
&& \frac{1}{r^2}{(1-\alpha^2 ) }- \frac{1}{r}{\alpha^2 ,}_{\, r }-
 \frac{\alpha^2 }{2r}\frac{{\delta^2 , }_{\, r }}{\delta^2 } \nonumber \\
&& =\kappa[\frac {F^2}{r^2} + V +\frac{1}{2}\rho -\frac{1}{2}p ] ,
\end{eqnarray}
which are consistent with those in the case of fermion stars \cite{TDL}.
The scalar field equation (21) reads
\begin{equation}
\alpha^2 {F,}_{\, rr}+\frac{\alpha^2 }{2}{F,}_{\, r}(\frac{{\delta^2
,}_{\,r}}{\delta^2} +2\frac{{\alpha^2 ,}_{\,r}}{\alpha^2 }
+\frac{4}{r})-\frac{2F}{r^2 }-\frac{\partial V}{\partial F}+Q F =0\, .
\end{equation}

If we assume that there exist regular solutions of the metric
components, scalar fields, and the energy density and pressure of
neutrinos to above equations (92)-(95) for small $r $, take series
expansions as
\begin{eqnarray}
\alpha^2 &=& \sum_{n\geq 0} a_n r^n \, ,\,\,\,\,\, \delta^2 =
\sum_{n\geq 0} b_n r^n \nonumber \\
F&=&\sum_{n\geq 0} \eta_n r^n \, ,\,\,\,\,\, \rho=\sum_{n\geq 0}\rho_n
r^n \, ,\,\,\,\,\, p=\sum_{n\geq 0} p_n r^n  ,
\end{eqnarray}
and substitute these expansions into Eqs.  (92)-(95), then we get the
following values for the coefficients.
\begin{eqnarray}
a_0 &=&1 \, ,\,\, a_1 =0 \, , \,\, a_2 = -\kappa[\frac{1}{3}\rho_0
+\frac{\lambda}{12}\eta^4 +\frac{1}{2}{\eta_1}^2 ] ,\nonumber \\
b_1& =&0 \, ,\,\,\,\,\, b_2 = b_0 \kappa [\frac{1}{2}\rho_0
+\frac{1}{2}p_0 +\frac{1}{2}{\eta_1}^2 ]\, ,\nonumber \\
\eta_0 &=&0 \, , \,\,\,\, \eta_2 =0 \, ,\,\,\,\, m_n \frac{v}{2 g_y}
Q_0 = \rho_0 -3 p_0 \, ...\,.
\end{eqnarray}
We thus have asymptotic solutions as
\begin{eqnarray}
F&=&\eta_1 r +{\cal O}(r^3 )\, , \nonumber \\
\alpha^2 & =&1-\kappa[\frac{1}{3}\rho_0 +\frac{\lambda}{12}\eta^4
+\frac{1}{2}{\eta_1}^2 ]r^2 +{\cal O}(r^3 ) ,\\
\delta^2 \alpha^2& =& \{ 1+\kappa[\frac{1}{6}\rho_0 +\frac{1}{2} p_0
-\frac{\lambda}{12}\eta^4 ]r^2 \} +{\cal O}(r^3 )\, ,\nonumber
\end{eqnarray}
where we have reparametrized the time as $\sqrt{b_0} t \rightarrow t$. 
These solutions are consistent with the results in fermionic stars
with a global monopole \cite{QGM}, even if they \cite{QGM} take $F
\simeq 0$ for small $r$. The value of $\eta_1$ is
determined by the matching condition that $F(r)$ should be continuous
at $r=r_c$, which is the upper bound  for small $r$. This gives us the relation,\begin{equation}
\eta_1 r_c =\eta.
\end{equation}
We might put $r_c \simeq 2M_{SBH}$ where $M_{SBH}$ is the mass of the
supermassive black hole at the center of our Galaxy \cite{BH}.

\section{Neutrino Clustering}

In Z-burst models to explain the continuation of the cosmic ray
spectrum beyond the GZK cutoff, Weiler \cite{Wei} and Fargion \cite{F}
assumed that the relic neutrinos cluster in galaxies a few times the
normal relic density, and found that the required flux of cosmic ray
neutrinos is larger than previously suggested.  Blanco-Pillado {\it et
al.} \cite{Bl} suggested that the relic neutrino density in the
clustering might be $10^{12} - 10^{14}$ times the mean relic density,
$n_M \simeq 54\, cm^{-3}$.  Independently in Ref. [5] a neutrino
cloud was considered as a sphere with no diffuse boundary such that
the relic neutrino density is given by $n_\nu (r) =n_R
\,\theta(R-r)+n_M \,\theta (r-R) $, where $n_R \simeq 10^{12} -10^{16}
\,cm^{-3}$ and $R\simeq 10^{14} - 10^{20} \, cm$.  This can moderate
the required incident flux of ultra high energy neutrinos.

From the results in the previous sections IV and V, we can have a
 cluster of neutrinos with a diffuse boundary such that the density of the
relic neutrinos is
\begin{equation}
n_\nu (r) =n_c \,\theta(r_c -r)+n_o (r) \,\theta (r-r_c) \theta (r_o -r), 
\end{equation}
where $n_c$ is the constant neutrino density for small $r \,(\leq r_c
\sim 2M_{SBH})$ as in the section V, and $n_o (r) =n_c
\frac{{r_c}^2}{r^2}$ is the neutrino density for $r_c < r \leq r_o$ as
in the section IV.  Here $r_o \simeq 10^{23} \,cm$(; the radius of our
Galaxy), $r_c \simeq 2M_{SBH} \simeq 10^{12} cm $(; the Schwarzschild
radius of a supermassive black hole at the center of the Galaxy), and
$n_c \geq 10^{3}(\frac{m_{eff}}{1\, eV})^3 \, cm^{-3}$ \cite{wax}.

\section{Summary and Discussions}
In the most general static spacetime with spherical symmetry, we
explicitly show that only $j=1/2$ fermions can satisfy Einstein's
equation, in the $\vec{\rho}\otimes\vec{\sigma}$ representation of
$\gamma$-matrices \cite{Dirac}. It is also possible to show the fact in
other representations of $\gamma$-matrices.  Considering neutrinos
coupled to $O(3)$ scalar fields $\vec{\Phi}$ via ${\vec{\Phi}}^2
\overline{\Psi}\Psi$-interaction term, we had a global monopole
solution \cite{Vil} of scalar fields and asymptotic solutions of the
metric components to Einstein's equation in Eqs.  (78) and (80), for
large $r$.  The asymptotic solution of one metric component $\delta^2
$ can be given in series of more general functions as $\delta^2 (r)
\simeq \sum_{n} {f_n}^2 (r) \, ln^n (\frac{r}{2M})$ with $f_n
(r\rightarrow \infty)=constant$, which do not change the asymptotic
behaviors of other solutions in this limit.

We adopted the Thomas-Fermi(TF) approximation for small $r$ expansions
of the energy density, pressure of neutrinos and so on.  Since only
$s$-waves are considered in the TF approximation, small $r$ expansions
in the section V are consistently related to asymptotic solutions
obtained for large $r$ in th esection IV.  More rigorous connection
between two asymptotic solutions can be made by numerical methods.  With the
simple assumption, Eq.  (100), about the radial dependence of the
relic density of neutrinos clustered in the galactic halo, further
studies of Z-burst models shall be possible.  For small $r$ solutions
in the section V to have astronomical meaning, the global $O(3)$
symmetry breaking scale $\eta$ and the scalar self-coupling constant
$\lambda$ in Eq.  (12) should be very small, which might be realized
in some Majoron models \cite{XHe}. In such a case the deficit angle
$1-c \equiv \kappa \eta^2 $ in Eq.  (78) is negligible \cite{Vil}.

\section*{ACKNOWLEDGEMENTS}

THL would like to thank School of Physics, The University of Melbourne
for the hospitality extended to him during the completion of this
article.  This work was supported by Korea Research Foundation
Grant(KRF-2001-015-DP0091).

\end{document}